\documentclass[12pt,osajnl2,preprint,showpacs,superscriptaddress]{revtex4}  
\usepackage[draft]{hyperref}
\usepackage{graphicx}
\usepackage{setspace}

\begin{document}

\setstretch{2}
\title{Electromagnetically-Induced-Transparency-Like Effect in the Degenerate Triple-Resonant
Optical Parametric Amplifier}
\author{Chenguang Ye}
\author{Jing Zhang$^{\dagger }$}
\affiliation{The State Key Laboratory of Quantum Optics and Quantum
Optics Devices, Institute of Opto-Electronics, Shanxi University,
Taiyuan 030006, P.R. China}

\begin{abstract}
We investigate experimentally the absorptive and dispersive
properties of  triple-resonant optical parametric amplifier OPA for
the degenerate subharmonic field. In the experiment, the subharmonic
field is utilized as the probe field and the harmonic wave as the
pump field. We demonstrate that EIT-like effect can be simulated in
the triple-resonant OPA when the cavity line-width for the harmonic
wave is narrower than that for the subharmonic field. However, this
phenomenon can not be observed in a double-resonant OPA. The narrow
transparency window appears in the reflected field. Especially, in
the measured dispersive spectra of triple-resonant OPA, a very steep
variation of the dispersive profile of the subharmonic field is
observed, which can result in a slow light as that observed in
atomic EIT medium.
\end{abstract}

\ocis{190.4410, 030.1670.}

\maketitle

Quantum coherence in atoms has led to a variety of novel phenomena.
In electromagnetically induced transparency (EIT), destructive
quantum interference introduced by a strong coupling laser cancels
the absorption from the ground state to coherent superposing upper
states \cite{one}. The observation of nonabsorbing resonance through
atomic coherence has led to novel concepts, such as lasing without
inversion \cite{two}. Since the EIT line-width can be made extremely
narrow, the resulting steep linear dispersion has been used to
reduce the velocity of light. This technique has been developed for
coherent optical information storage and to freeze light
\cite{three,three1,three2,four}.

Since EIT results from destructive quantum interference, it has been
recognized that similar coherence and interference effects can also
occur in classical systems, such as plasma \cite {five,five1},
coupled optical resonators \cite{six,six2,six3,six4,six1},
mechanical or electric oscillators \cite{seven}. Particularly, the
phenomenology of the EIT and the dynamic Stark effect are studied
theoretically in a dissipative system composed by two coupled
oscillators using quantum optics model in Ref. \cite{eight}. The
optical parametric amplifier (OPA) is one of the most important
nonlinear and quantum optical devices. Recently, we studied the
coherence phenomena in the phase-sensitive degenerate OPA inside a
cavity \cite{nine,ten}. This phenomenon results from the
interference between the harmonic pump field and the subharmonic
seed field in OPA. The destructive and constructive interferences
correspond to optical parametric deamplification and amplification
respectively. The absorptive response of an optical cavity for the
probe field is changed by optical parametric interaction in the
cavity. Agarwal \cite{eleven} generalized our observation of
interferences in the classical domain \cite{nine} to in the quantum
domain and studied the interferences in the quantum fluctuations of
the output fields from a parametric amplifier when the cavity is
driven by a quantized field at the signal frequency.

In the previous works \cite{nine,ten}, we studied the absorptive and
dispersive properties of degenerate double-resonant OPA where only
the degenerate subharmonic field resonated inside the cavity but the
pump field did not. Although in the experiment with double-resonant
OPA the mode splitting in the transmission spectra \cite{nine} and
the M shape in the reflection spectra were observed experimentally
\cite{ten}, the shape of the phase shift of the reflected field was
unchanged \cite{ten}. Further, we investigated the triple-resonant
OPA theoretically, in which both the subharmonic and harmonic wave
ares simultaneously resonate inside the cavity \cite{ten}. We found
that the EIT-like effect can be simulated with a triple-resonant
OPA. In this Letter, we present the results of the experimental
investigation on the absorptive and dispersive properties of a
triple-resonant OPA for the subharmonic field. The frequency and
polarization of signal and idler fields in the OPA are totally
degenerate. We demonstrate that EIT-like effect can be simulated in
OPA when the cavity line-width for the harmonic wave is narrower
than that for the subharmonic field. A narrow transparency window
appears in the reflected field. Especially, we measured the
dispersive spectra of triple-resonant OPA and a very steep variation
in the dispersive profile was observed.

\begin{figure}
\centerline{
\includegraphics[width=6.3in]{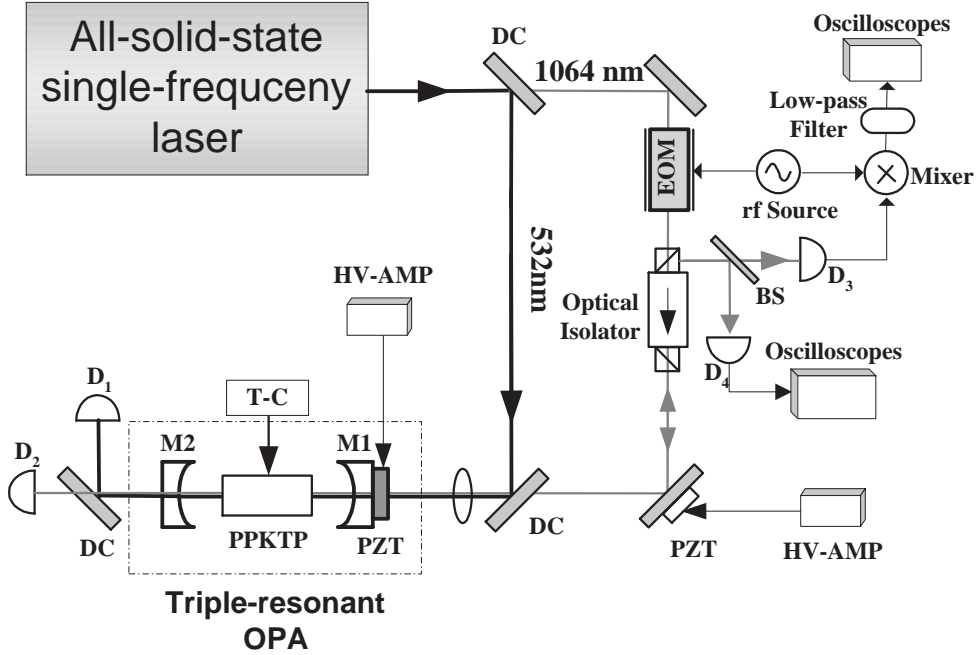}
} \vspace{0.1in}
\setstretch{2} \caption{ Schematic of the experimental setup. DC:
dichroic mirror; $\lambda /2$: half-wave plate; $D_{1},D_{2}, D_{3
}, D_{4}$: detectors; T-C: temperature controller, HV-AMP: high
voltage amplifier; PZT: piezoelectric transducer; EOM:
electric-optic modulator. \label{Fig3} }
\end{figure}

The experimental setup is shown schematically in Fig. 1, which is
similar to that used in our previous work \cite{ten}. A diode-pumped
intra-cavity frequency-doubled continuous-wave ring
Nd:YVO$_{\text{4}}$/KTP single-frequency laser severs as the light
sources of the pump wave (the second-harmonic wave at 532 nm) and
the probe wave (the subharmonic wave at 1064 nm) for OPA. We
actively control the relative phase between the subharmonic and the
pump field by adjusting the phase of the subharmonic beam with a
mirror mounted upon a piezoelectric transducer (PZT). Both beams are
combined together by a dichroic mirror and injected into the OPA
cavity. OPA consists of type I periodically poled KTiOPO$_4$ (PPKTP)
crystal (12 mm long) and two external mirrors separated by 61 mm.
Two end faces of crystal are polished and coated with an
antireflection for both wavelengths of 532 nm and 1064 nm. The
crystal is mounted in a copper block, whose temperature is actively
controlled at millidegrees kelvin level around the phase-match
temperature for optical parametric process (31.3${{}^{\circ }}$C).
The input coupler $M1$ is a 30 mm radius-of-curvature mirror with a
power reflectivity $99\%$ for 1064 nm in the concave and $97\%$ for
532nm, which is mounted upon a PZT for adjusting the length os the
optical cavity. The back mirror $M2$ is a 30-mm radius-of-curvature
mirror with a reflectivity $93\%$ for 1064 nm and a high
reflectivity coefficient for 532 nm in the concave. This structure
forms the triple-resonant OPA, in which the pump wave and the
frequency-degenerated idler and signal waves are all resonating
inside the same cavity. This is distinctly different with our
previous setup. The cavity is a under-coupled resonator for the
subharmonic field due to $\gamma _{in}<\gamma _c+\gamma _l$, where
$\gamma _{l}$,$\gamma _c$ and $\gamma _{in}$ are the decay rates of
subharmonic field resulting from internal losses, input mirror and
back mirror respectively. The cavity line-width for the harmonic
wave is narrower than for the subharmonic field because the total
damping of the pump field is much less than that of the subharmonic
field ($\gamma _b\ll \gamma )$. Due to the high nonlinear
coefficient of PPKTP, the measured threshold power is $P_{th}=90$
$mW$. In order to measure the phase of the reflected beam, we employ
the Pound-Drever-Hall method (or Frequency-modulation spectrum)
\cite
{thirty,thirty-one,thirty-two,thirty-three,thirty-four,thirty-five},
which provides a way to indirectly measure the phase. The frequency
of the injected subharmonic field is modulated with an
electric-optic modulator (EOM), driven by a local oscillator with a
radio frequency of $100$ $MHz$. Here the depth of frequency
modulation is very small, thus we only can observe two sidebands.
The reflected beam is picked off with an optical isolator and send
into a photodetector $D_{3}$ , whose output is compared with the
local oscillator's signal via mixer. A low-pass filter on the output
of the mixer extracts the low frequency signal, which is called the
error signal. When the frequency of the injected subharmonic field
is near resonance and the modulation frequency is high enough, the
sidebands are out of the resonance and the error signal is the
imaginary part of the reflected field. In our scheme, the cavity is
an under-coupled resonator, whose dispersive response results in
fast light \cite{six1}. When an under-coupled resonator is far from
critically-coupled, the imaginary part of the reflected field
approximates to the phase shift.

\begin{figure}
\centerline{
\includegraphics[width=6.8in]{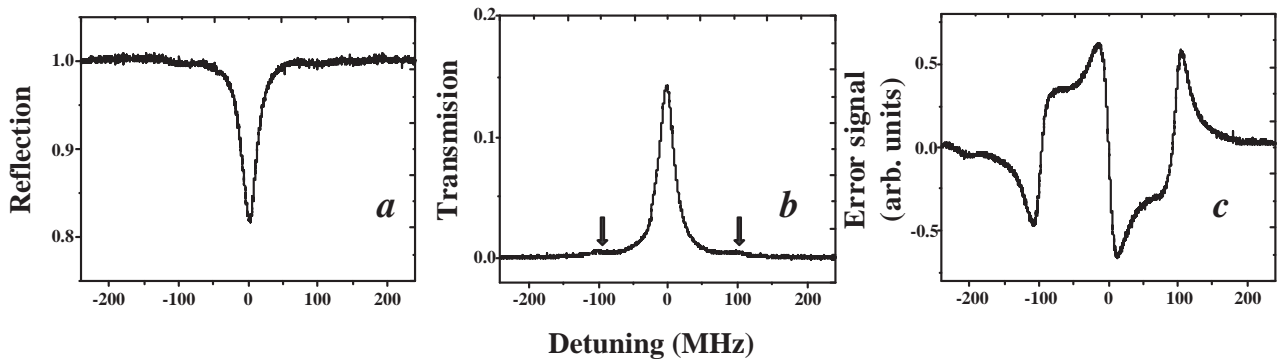}
} \vspace{0.1in}
\setstretch{2} \caption{ The spectra of the subharmonic reflection,
transmission and phase shift of the reflected field as a function of
the cavity detuning when the pump field is blocked. \label{Fig2} }
\end{figure}

First, we measured the spectra of the subharmonic reflection,
transmission and phase shift of the reflection as the functions of
the cavity detuning at the case without the existence of the pump
field, which are shown in Fig.2. Two small sideband modes generated
by frequency modulation can be seen in the transmission spectrum
(See arrows in Fig.2 b). Then we hold the frequencies of the
injected subharmonic and the pump field at $\omega _p=2\omega $, and
the phase difference $\varphi =\pi /2$ (i.e. OPA at the state of
deamplification), and mode two optical fields resonate
simultaneously in the OPA by adjusting the cavity length and the
temperature of PPKTP. The cavity length is scanned around the
resonant point. In Fig.3, the curves e, f and g show the spectra of
the subharmonic reflection, transmission and phase shift of the
reflection respectively with the pump power below the threshold
$P=0.15P_{th}$, and the curves h, j and k with $P=0.3P_{th}$. We can
see that the narrow transparency window appears in the reflection
spectrum and is accompanied by a very steep variation of the
dispersive profile whose dispersive response can result in slow
light as that in EIT medium. This experimental result is in good
agreement with the theoretical predictions in Ref.\cite{ten}.

\begin{figure}
\centerline{
\includegraphics[width=6.8in]{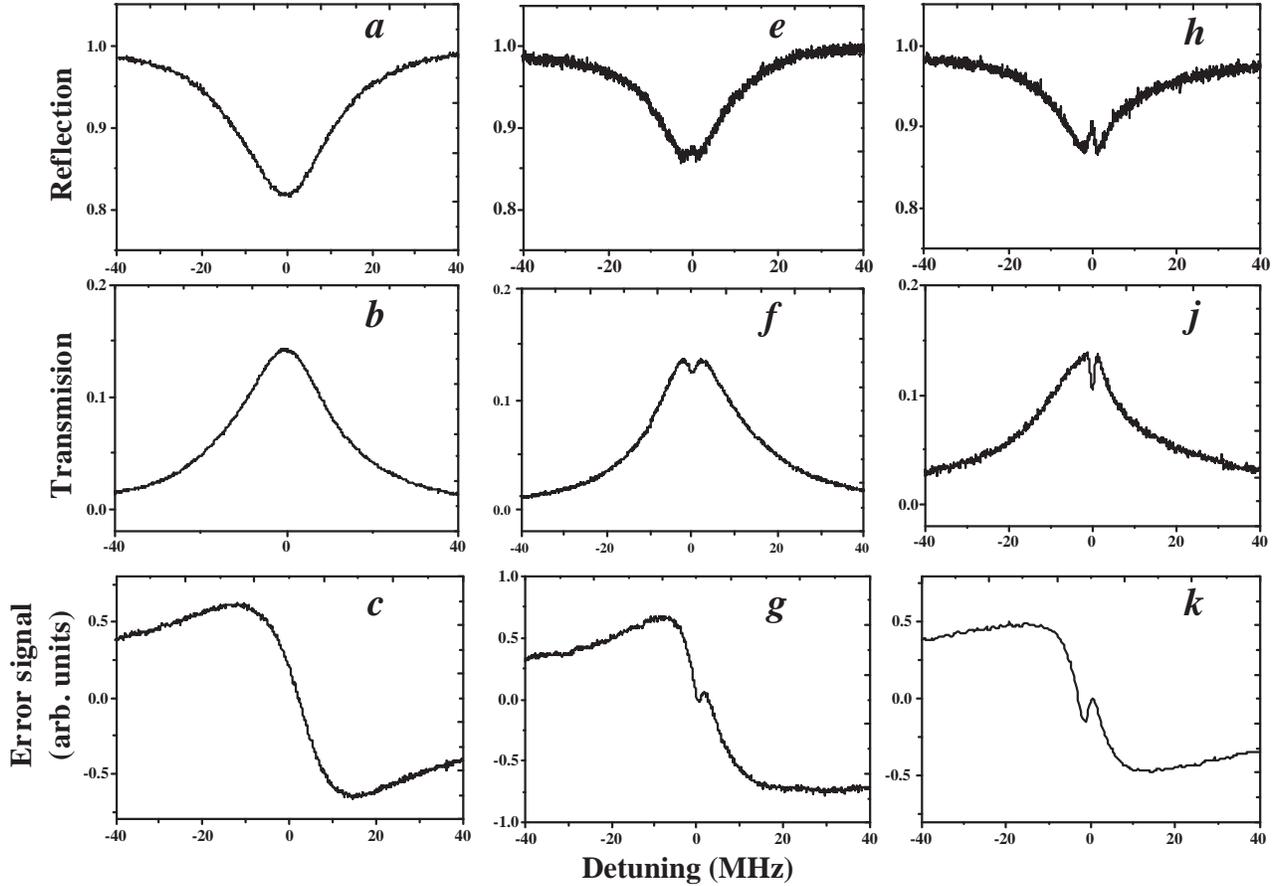}
} \vspace{0.1in}
\setstretch{2} \caption{ The spectra of the subharmonic reflection,
transmission and phase shift of the reflected field as a function of
the cavity detuning for different pump powers. a,b and c correspond
to Fig.2 with narrow frequency detuning. e, f, and g at the pump
power $0.15P_{th}$. h, j and k at $0.3P_{th}$. \label{Fig3} }
\end{figure}

The observed phenomenon is easily understood: since the narrow
absorption and dispersion of the pump field are introduced into the
subharmonic field via the parametric interaction in triple-resonant
OPA, the reflected field of the subharmonic field presents the
EIT-like effect \cite{ten}. For the case of EIT in atomic systems,
the analogous condition of $\gamma _b\ll \gamma $ is obtained when a
pair of lower energy states have long lifetime. Note that we did not
observe the phenomena under higher pump power as that predicted
theoretically in \cite{ten}. There are two main reasons. One is that
the OPA will produce a large quantum fluctuation on the subharmonic
field under higher pump power. The other is that the pump field had
the strong thermal effect at the high pump power, which would
influence the constructed triple-resonance. When scanning the cavity
length, the thermo-induced effect for the pump field was observed as
shown in Fig.4, from which we can see, it seems like the Kerr
nonlinear effect. The transmission profile of the pump field appears
large asymmetry at higher pump power when cavity length is scanned
from shorter to longer and when the opposite. Thus it is difficult
to measure the unusual line shape in the reflection spectrum and the
steep variation of the dispersive profile for triple-resonant OPA in
the case of the strong parametric coupling between two fields.

\begin{figure}
\centerline{
\includegraphics[width=6.8in]{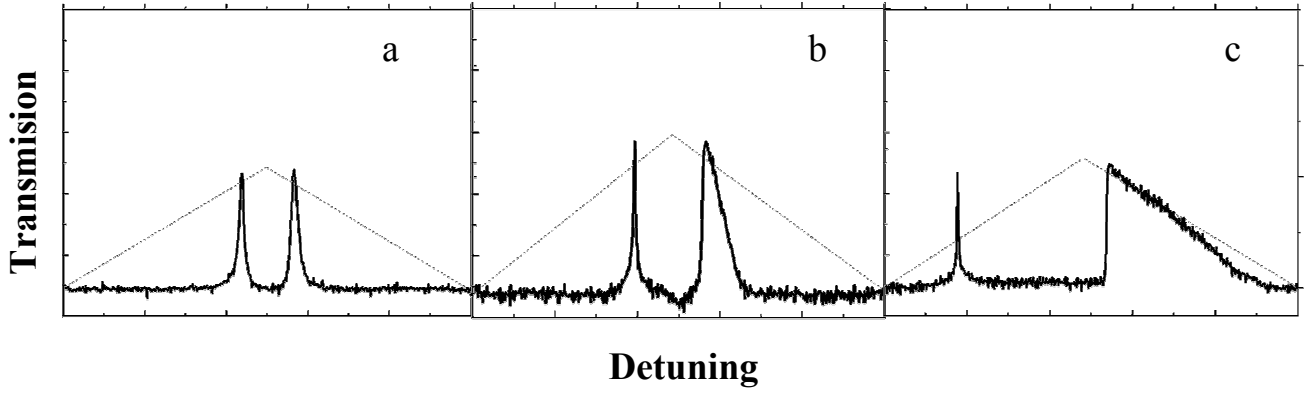}
} \vspace{0.1in}
\setstretch{2} \caption{ The spectra of transmission of the pump
field as a function of the cavity detuning for different pump
powers: a at $0.2P_{th}$; b at $0.5P_{th}$; c at $0.9P_{th}$. The
dot line represents the voltage on PZT, which denotes the cavity
length scanned from shorter to longer, then from longer to shorter.
\label{Fig4} }
\end{figure}

In conclusion, we have investigated experimentally the absorptive
and dispersive response of the reflected subharmonic field from a
triple-resonant phase-sensitive OPA. It is demonstrated that an
EIT-like effect is simulated when the cavity line-width for the
harmonic wave is narrower than for the subharmonic field. The narrow
transparency window appears in the reflected field and is
accompanied by a very steep variation of the dispersive profile.
This novel system accompanying parametric gain for the probe signal
will be important for practical optical and photonic applications
such as in optical filters, delay lines, and closely relate to the
coherent phenomenon of EIT as that predicted for quantum systems.

$^{\dagger} $Corresponding author's email address: jzhang74@yahoo.com,
jzhang74@sxu.edu.cn

\smallskip \acknowledgments
J. Zhang thanks K. Peng and C. Xie for the helpful discussions. This
research was supported in part by NSFC for Distinguished Young
Scholars (Grant No. 10725416), National Basic Research Program of
China (Grant No. 2006CB921101), National Natural Science Foundation
of China (Grant No. 60678029).

REFERENCES


\begin{thebibliography}{99}

\bibitem{one}  S. E. Harris, Phys. Today {\bf 50}, 37 (1997); J. P.
Marangos, J. Mod. Opt. {\bf 45}, 471 (1998).

\bibitem{two}  A. S. Zibrov, M. D. Lukin, D. E. Nikonov, L. Hollberg, M. O. Scully, V. L. Velichansky, and H. G. Robinson, Phys. Rev. Lett. {\bf 75}, 1499
(1995).

\bibitem{three} L. V. Hau, S.E. Harris, Z. Dutton and C.H. Behroozi, Nature (London) {\bf 397}, 594
(1999).
\bibitem{three1} M. M. Kash, V. A. Sautenkov, A. S. Zibrov, L. Hollberg, G. R. Welch,
M. D. Lukin, Y. Rostovtsev, E. S. Fry, and M. O. Scully, Phys. Rev.
Lett. {\bf 82}, 5229 (1999).
\bibitem{three2} D. Budker, D. F. Kimball, S. M.
Rochester, and V. V. Yashchuk, Phys. Rev. Lett. {\bf 83}, 1767
(1999).

\bibitem{four} C. Liu, Z. Dutton, C. Behroozi and L. V. Hau, Nature (London) {\bf 409}, 490 (2001);
D. F. Phillips, A. Fleischhauer, A. Mair, and R. L. Walsworth, Phys.
Rev. Lett. {\bf 86}, 783 (2001).


\bibitem{five}  S.E. Harris, Phys. Rev. Lett. {\bf 77}, 5357 (1996).

\bibitem{five1} A.G. Litvak and M.D. Tokman, Phys. Rev. Lett. {\bf 88}, 095003
(2002); G. Shvets and J.S Wurtele, Phys. Rev. Lett. {\bf 89}, 115003
(2002).

\bibitem{six} T. Opatrny, and D. G. Welsch, Phys. Rev. A {\bf64}, 023805 (2001).

\bibitem{six2} D. D. Smith, H. Chang, K. A. Fuller, A. T. Rosenberger, and R. W. Boyd,
Phys. Rev. A {\bf 69}, 063804 (2004); L. Maleki, A. B. Matsko, A. A.
Savchenkov, and V. S. Ilchenko, Opt. Lett. {\bf 29}, 626 (2004).

\bibitem{six3}M. F. Yanik, W. Suh, Z. Wang, and S. Fan, Phys. Rev. Lett. {\bf 93},
233903 (2004).

\bibitem{six4} A. Naweed, G. Farca, S. I. Shopova, and A. T.
Rosenberger, Phys. Rev. A {\bf71}, 043804 (2005); Q. Xu, S. Sandhu,
M. L. Povinelli, J. Shakya, S. Fan, and M. Lipson, Phys. Rev.Lett.
{\bf 96}, 123901 (2006);

\bibitem{six1} D.D. Smith, H. Chang, J. Mod. Opt. {\bf 51}, 2503 (2004);

\bibitem{seven}  P.R. Hemmer and M.G. Prentiss, J. Opt. Soc. Am. B {\bf 5}, 1613 (1988);
C. L. Garrido Alzar, M. A. G. Martinez, P. Nussenzveig, Am. J. Phys.
{\bf 70}, 37 (2002).


\bibitem{eight}  M. A. de Ponte, C. J. Villas-Boas, R. M. Serra, and M. H. Y. Moussa, e-print quant-ph/0411087.

\bibitem{nine} H. Ma, C. Ye, D. Wei, and J. Zhang, Phys. Rev. Lett. {\bf95}, 233601 (2005).

\bibitem{ten} C. Ye and J. Zhang, Phys. Rev. A {\bf73}, 023818(2006).

\bibitem{eleven} G. S. Agarwal, Phys. Rev. Lett. {\bf97}, 023601 (2006).

\bibitem{thirty} R. W. P. Drever, J. L. Hall, F. V. Kowalski, J. Hough,
G. M. Ford, A. J. Munley, and H. Ward, Appl. Phys. B {\bf31}, 97
(1983).

\bibitem{thirty-one} G. C. Bjorklund, Opt. Lett. {\bf5}, 15 (1980).

\bibitem{thirty-two} G. C. Bjorklund, M.  D. Levenson, W. Lenth, C. Ortiz, Appl. Phys. B {\bf32}, 145 (1983).

\bibitem{thirty-three} A. Schenzle, R. G. DeVoe, and R. G. Brewer, Phys. Rev. A {\bf25}, 2606 (1982).

\bibitem{thirty-four} J. Zhang, D. Wei, C. Xie and K. Peng, Opt. Express {\bf11}, 1338 (2003).

\bibitem{thirty-five} E. D. Black, Am. J. Phys. {\bf69}, 79 (2001).

\end{thebibliography}
\end{document}